# Resource Allocation for Energy-Efficient 3-Way Relay Channels

Bho Matthiesen, *Student Member, IEEE,* Alessio Zappone, *Member, IEEE,*

Eduard A. Jorswieck, *Senior Member, IEEE,*

## Abstract

Throughput and energy efficiency in 3-way relay channels are studied in this paper. Unlike previous contributions, we consider a circular message exchange. First, an outer bound and achievable sum rate expressions for different relaying protocols are derived for 3-way relay channels. The sum capacity is characterized for certain SNR regimes. Next, leveraging the derived achievable sum rate expressions, cooperative and competitive maximization of the energy efficiency are considered. For the cooperative case, both low-complexity and globally optimal algorithms for joint power allocation at the users and at the relay are designed so as to maximize the system global energy efficiency. For the competitive case, a game theoretic approach is taken, and it is shown that the best response dynamics is guaranteed to converge to a Nash equilibrium. A power consumption model for mmWave board-to-board communications is developed, and numerical results are provided to corroborate and provide insight on the theoretical findings.

## I. Introduction

Relays are fundamental building blocks of wireless networks. Deploying relays in areas affected by significant shadowing such as tunnels or the inside of buildings, or in areas that are far away

Part of this work has been published in [1].

The work of Bho Matthiesen and Eduard Jorswieck is supported by the German Research Foundation (DFG) in the Collaborative Research Center 912 "Highly Adaptive Energy-Efficient Computing." The work of Alessio Zappone has been funded by the German Research Foundation (DFG) project CEMRIN, under grant ZA 747/1-2.

The authors are with the Chair for Communications Theory, Communications Laboratory, Technische Universität Dresden, Dresden, Germany (e-mail: bho.matthiesen@tu-dresden.de, alessio.zappone@tu-dresden.de, eduard.jorswieck@tu-dresden.de).

from the transmitter, allows to extend cell-coverage and increase the network's reliability and throughput. Thus, the study of relay channels is essential in understanding the capacity limits of modern networks and for the development of novel communication schemes. One recently-proposed channel model for relay networks is the multi-way relay channel (MWRC), which models clustered communication over a relay, where the terminals in each cluster exchange information among each other with the help of a relay. Such a model applies to many communication architectures like the communication of several ground stations over a satellite, or wireless board-to-board communication in highly adaptive computing [2] where multiple chips exchange data with the help of another chip acting as relay. The MWRC was first introduced in [3], where all users in the cluster send a message and are interested in decoding the messages of all other users in the cluster. In [4] the common-rate capacity of the additive white Gaussian noise (AWGN) MWRC with full message exchange is derived and it is shown that for three and more users this capacity is achieved by decode-and-forward (DF) for signal-to-noise ratios (SNRs) below $0\,\text{dB}$ and compute-and-forward otherwise. The same authors present in [5] the capacity region of the finite field MWRC. In [6] a constant gap approximation of the capacity region of the Gaussian 3-user MWRC with full message exchange is derived. In contrast to most other works, the authors consider private messages instead of common messages, i.e., each user transmits distinct messages to the other users instead of a common one.

Besides throughput, another key performance metric in modern and future 5G wireless networks is energy efficiency (EE), which is steadily gaining momentum due to green and sustainable growth concerns. The need for EE is even stronger for battery-powered terminals, in order to extend their lifetimes. From a mathematical standpoint, one well-established definition of the EE of a communication system is the ratio between the system capacity or achievable rate and the total consumed power [7], [8]. With this definition, the EE is measured in bit/Joule (or bit/Hz/Joule when the rate is given in bit/s/Hz), thus naturally representing the efficiency with which the available energy is used to transmit information. Previous results on EE in relay systems mainly focus on one-way amplify-and-forward (AF) or DF schemes and do not consider the MWRC. In [9] the optimal placement of relays in cellular networks is investigated and is seen to provide power-saving gains. In [10] instead, cooperative approaches aimed at weighted sum-power minimization with fairness and rate constraints are devised. In [11] a pricing-based approach is employed to come up with energy-saving power control algorithms. There, the EE is defined as the difference of the achievable rate and the transmit power scaled by a price. Instead, [12] considers the bit/Joule definition of EE and devises energy-efficient power control algorithms in interference networks. A cooperative approach is considered in [13], where a multiple-input multiple-output (MIMO) AF relay-assisted system is considered and the source and relay precoding matrices are allocated so as to maximize the global EE, for different channel state information (CSI) assumptions [14]. Power control in one-way and two-way relay channels with AF is studied in [15].

In this paper a 3-way relay channel is considered and both throughput and energy efficiency are analyzed and optimized. In contrast to most other works on MWRCs, we focus on a partial message exchange where each message is only intended for one receiver and, also, not every user sends a message to each other user. This might grant higher achievable rates provided one can deal with interference at the receivers which makes the analysis more involved.

In such a scenario, we make the following contributions:

1) We derive achievable sum rate expressions for the AF, DF, noisy network coding (NNC) relaying protocols, and an outer bound on the sum capacity. The sum capacity is characterized for certain channel configurations and shown to be achieved by DF relaying. For completely symmetric channels, we show that the sum capacity is achieved by DF for SNRs below $8.1\,\text{dB}$, and within a constant gap by NNC and AF.

2) Based on the derived achievable sum rate expressions, cooperative power control algorithms to maximize the system global energy efficiency (GEE) are provided. If DF is used at the relay, the global optimum of the system GEE is found in polynomial time by means of fractional programming theory. For all other relaying schemes, leveraging fractional programming and monotonic optimization theory, two algorithms are provided to compute both a local solution with low complexity and the global optimum with higher complexity. Numerical evidence suggests that both algorithms actually achieve the global optimum.

3) Energy-efficient power control is also investigated in a competitive scenario formulating the problem as a non-cooperative game. The existence of a Nash equilibrium (NE) is shown and a best response dynamics (BRD)-based algorithm is provided which is guaranteed to always converge to a NE point.

4) The performance of the proposed algorithms and of the different relaying schemes are thoroughly compared by numerical simulations. A realistic power consumption model for a mmWave board-to-board communication system operating at $200\,\text{GHz}$ is developed. It is shown that, although not achieving the highest sum rate, AF relaying is more energy-efficient than all other schemes due to its low hardware complexity.

We define the function $\text{C}(x) = \log_2(1+x)$ for $x \geq 0$. $f'(x)$ and $f''(x)$ denote the first and second derivative of $f$ as functions of $x$, respectively. $\log(\cdot)$ denotes natural logarithms.

## II. System Model

We consider a 3-user single-input single-output (SISO) MWRC relay-assisted system in which the three users communicate with each other via a relay node. A symmetric scenario is considered with a circular (i.e. partial) message exchange, to be described in detail later. We assume full-duplex transmission and consider a scenario in which no direct user-to-user link is available. The users are denoted as node 1 to 3 and the relay as node $R$. We define the set of all users as $\mathcal{K} = \{1, 2, 3\}$ and the set of all nodes as $\mathcal{K}_R = \mathcal{K} \cup \{R\}$.

The considered communication system can be decomposed in a multiple-access channel (MAC), which carries the superposition of the signals from the three users to the relay, and a broadcast channel (BC), which conveys the signal from the relay to the users. Specifically, the signal received by the relay is given by $Y_R = \sum_{k \in \mathcal{K}} X_k + Z_R$, with $X_k$ the channel input at node $k \in \mathcal{K}_R$ and $Z_R$ the zero mean circularly symmetric complex Gaussian noise with power $N_R$. The relay processes $Y_R$ to produce a new message $X_R$, which is then broadcasted to the users. Then, for all $k \in \mathcal{K}_R$, the signal received at user $k$ is given by $Y_k = X_R + Z_k$, with $Z_k$ the zero mean circularly symmetric complex Gaussian noise with power $N_S$. All noise variables are mutually independent and the channel inputs are independent and identically distributed (i.i.d.) over time. All channel inputs have zero mean and an average power constraint $\mathbb{E}\,|X_R|^2 \leq P_R$ and $\mathbb{E}\,|X_k|^2 \leq P_S$, for $k \in \mathcal{K}$. The rule which defines how $X_R$ is produced from $Y_R$ depends on the particular relay protocol. In this paper, we will consider DF, NNC, and AF, both with treating interference as noise (IAN) and with simultaneous non-unique decoding (SND) at the receivers. To conclude this section, we describe in detail the considered circular message exchange.

We consider the partial message exchange illustrated in Fig. 1. It has two defining properties:

1) Each user has a message to transmit which is intended for at least one other user.
2) Each user desires at most one message.

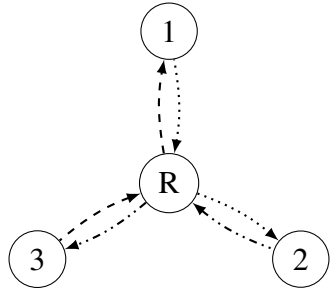

Fig. 1. Illustration of the system model where node 0 is the relay and nodes 1 to 3 are the users. Messages travel along the different line styles.

From these two properties it follows immediately that each message is only required at one other user. We denote the message of user $k$, $k \in \mathcal{K} = \{1,2,3\}$, as $m_k$ and the node receiving it is denoted by $q(k)$. Furthermore, the user not interested in $m_k$ is denoted as $l(k)$. Also, it follows from the properties listed above, that user $k$, $k \in \mathcal{K}$, desires the message sent by user $l(k)$. Further, since $m_k$ and $m_{l(k)}$ are to be decoded by users $q(k)$ and $k$, respectively, $m_{q(k)}$ is the desired message at node $l(k)$. Without loss of generality, we assume a clockwise message exchange with $q(k)$ and $l(k)$ defined as follows:

$$q(k) = \begin{cases} 2 & \text{if } k = 1 \\ 3 & \text{if } k = 2 \\ 1 & \text{if } k = 3 \end{cases}, \qquad l(k) = \begin{cases} 3 & \text{if } k = 1 \\ 1 & \text{if } k = 2 \\ 2 & \text{if } k = 3 \end{cases}.$$

A $(2^{nR_1}, 2^{nR_2}, 2^{nR_3}, n)$ code for the 3-user MWRC consists of three message sets $\mathcal{M}_k = [1 : 2^{nR_k}]$, one for each user $k \in \mathcal{K}$, three encoders, where encoder $k \in \mathcal{K}$ assigns a symbol $x_{ki}(m_k, y_k^{i-1})$ to each message $m_k \in \mathcal{M}_k$ and received sequence $y_k^{i-1}$ for $i \in [1 : n]$, a relay encoder that assigns a symbol $x_{0i}(y_0^{i-1})$ to every past received sequence $y_0^{i-1}$ for $i \in [1 : n]$, and three decoders, where decoder $q(k) \in \mathcal{K}$ assigns an estimate $\hat{m}_k \in \mathcal{M}_k$ or an error message $e$ to each pair $(m_{q(k)}, y_{q(k)}^n)$.

We assume that the message triple $(M_1, M_2, M_3)$ is uniformly distributed over $\mathcal{M}_1 \times \mathcal{M}_2 \times \mathcal{M}_3$. The average probability of error is defined as $P_e^{(n)} = \Pr\left\{\hat{M}_k \neq M_k \text{ for some } k \in \mathcal{K}\right\}$.

A rate triple $(R_1, R_2, R_3)$ is said to be achievable if there exists a sequence of $(2^{nR_1}, 2^{nR_2}, 2^{nR_3}, n)$ codes such that $\lim_{n \to \infty} P_e^{(n)} = 0$. The capacity region of the 3-user MWRC is the closure of the set of achievable rates. The sum rate is defined as $R_\Sigma = \max\{R_1 + R_2 + R_3 : (R_1, R_2, R_3) \in \mathcal{R}\}$, where $\mathcal{R}$ is an achievable rate region. Whenever $\mathcal{R}$ is the capacity region, we call $R_\Sigma$ the sum capacity $C_\Sigma$.

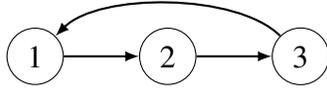

Fig. 2. Message side information graph as defined in [16] for the considered system model. A vertex $k$ represents the receiver of message $M_k$ and is not to be confused with "user $k$" as referred to in the system model. There is a directed edge from $k$ to $j$ if and only if the receiver of $M_k$ knows $M_j$ a priori. For example, vertex 1 receives message $M_1$ and knows message $M_{q(1)}$ a priori.

## III. BOUNDS ON THE SUM CAPACITY

We start our treatment of the symmetric 3-user MWRC by deriving an upper bound on the sum capacity and then continue with several inner bounds.

### A. Outer Bound

This outer bound consists of the cut set bound in the uplink and a downlink bound [16] that takes the side information at the receivers into account.

*Lemma 1:* The sum capacity of the symmetric 3-user MWRC is upper bounded as

$$C_\Sigma \leq \min\left\{\frac{3}{2}\,\mathrm{C}\left(\frac{P_R}{N_S}\right),\, 3\,\mathrm{C}\left(\frac{P_S}{N_R}\right)\right\}. \tag{1}$$

*Proof:* Assume the relay knows the messages $M_1$, $M_2$, and $M_3$ a priori. Then we are looking at a 3-user BC with receiver message side information. An outer bound to this channel is given in [16, Theorem 4] where a node $k$ receives a message $M_k$ and knows a subset $\mathcal{M}_k$ of messages a priori as side information.

In our model, node $q(k)$ receives message $M_k$ and knows message $M_{q(k)}$ a priori, i.e., $\mathcal{M}_k = \{M_{q(k)}\}$. The resulting message side information graph as defined in [16] is shown in Fig. 2. Please note that vertex $k$ in Fig. 2 represents the receiver of message $M_k$ and, thus, corresponds to user $q(k)$. The message side information graph is cyclic, but it is easily seen that all induced subgraphs are acyclic. Hence, from [16, Theorem 4] we have that every achievable rate tuple $(R_1, R_2, R_3)$ must satisfy $R_i + R_j \leq \mathrm{C}\left(\frac{P_R}{N_S}\right)$, for $i \neq j$ and $i, j \in \mathcal{K}$. Applying the simplex algorithm (see, e.g., [17]) analytically, we can show that $R_\Sigma \leq \frac{3}{2}\,\mathrm{C}\left(\frac{P_R}{N_S}\right)$.

For the second bound, apply the cut set bound [18, Theorem 15.10.1] to the cuts $\mathcal{S} = \{k\}$, $k \in \mathcal{K}$. This results in $R_k \leq \mathrm{C}\left(\frac{P_S}{N_R}\right)$, for $k \in \mathcal{K}$, and, thus, $R_\Sigma \leq 3\,\mathrm{C}\left(\frac{P_S}{N_R}\right)$.  ∎

## *B. Amplify-and-Forward*

We first consider AF relaying where the relay scales the observed signal by a positive constant and broadcasts it back to the users. The transmitted symbol at the relay is $X_R = \alpha Y_R$, where $\alpha$ is a normalization factor chosen such that the transmit power constraint at the relay is met, i.e., $\alpha = \sqrt{P'_R / \left(\sum_{k \in \mathcal{K}} P'_k + N_R\right)}$, where $P'_k$, $k \in \mathcal{K}_R$, is the actual transmit power of node $k$ satisfying the average power constraints. The receiver first removes its self-interference from the received signal[1] and then decodes for its desired message while treating the remaining interference as noise. We split the transmission into three equal length blocks, i.e., time slots, and switch off user $i$ in time slot $i$, $i = 1, 2, 3$. This reduces interference and allows for higher transmission powers in the other two time slots while still meeting the average power constraint.

*Lemma 2:* In the 3-user MWRC, the sum rate

$$R_\Sigma^{AF-IAN} = C\left(\frac{3P_S P_R}{N_R P_R + 3P_S N_S + N_S N_R}\right) \quad (2)$$

is achievable with AF relaying and IAN at the receivers.

*Proof:* The received signal of user $q(k)$ is $Y_{q(k)} = \sum_{j \in \mathcal{K}} \alpha X_j + \alpha Z_R + Z_{q(k)}$. Since $X_{q(k)}$ is known at receiver $q(k)$, it can be removed from the received signal. Thus, $\tilde{Y}_{q(k)} = Y_{q(k)} - \alpha X_{q(k)} = \alpha X_k + \alpha X_{l(k)} + \alpha Z_R + Z_{q(k)}$. Treating the interfering symbol $X_{l(k)}$ as additional noise, the maximum achievable rate $R_k$ is $C\left(\frac{\alpha^2 P'_k}{\alpha^2 P'_{l(k)} + \alpha^2 N_R + N_S}\right)$. Plugging in the expression for $\alpha$ we have

$$R_k \leq C\left(\frac{P'_R P'_k}{P'_R P'_{l(k)} + P'_R N_R + \left(\sum_{j \in \mathcal{K}} P'_j + N_R\right) N_S}\right), \quad (3)$$

for all $k \in \mathcal{K}$.

By computing its first-order derivative, it is shown that the right-hand side (RHS) of (3) is monotonically increasing in $P'_R$. Thus, $P'_R = P_R$ is optimal.

---

[1]Due to the static channels and long transmission times, we can assume without loss of generality that $\alpha$ is known at all nodes.

Next, we split the transmission into three equal length blocks, i.e. time slots, and switch off user $i$ in time slot $i$, i.e., $P_i'^i = 0$,[2] $i = 1, 2, 3$. Then, the achievable sum rate in time slot $i$ is

$$R_\Sigma^i = \frac{1}{3}\left(R_{q(i)} + R_{l(i)}\right) = \frac{1}{3}\,\mathrm{C}\left(\frac{P_R P_{q(i)}'^i}{P_R N_R + \left(P_{q(i)}'^i + P_{l(i)}'^i + N_R\right)N_S}\right) \\ + \frac{1}{3}\,\mathrm{C}\left(\frac{P_R P_{l(i)}'^i}{P_R P_{q(i)}'^i + P_R N_R + \left(P_{q(i)}'^i + P_{l(i)}'^i + N_R\right)N_S}\right). \quad (4)$$

Since each user transmits only in two out of three time slots the average power constraint is met as long as $\sum_{j\in\mathcal{K}\setminus\{k\}} P_k'^j \leq 3P_S$ for all $k \in \mathcal{K}$. From the first order derivatives of (4) it is obtained that (4) is monotonically increasing in $P_{q(i)}'^i$ and $P_{l(i)}'^i$. Thus, this constraint should be met with equality.

We set $P_1'^2 = 3P_S - p$, $P_1'^3 = p$, $P_2'^3 = 3P_S - p$, $P_2'^1 = p$, $P_3'^1 = 3P_S - p$, $P_3'^2 = p$, for some $p \in [0, 3P_S]$ in (4) and get

$$\frac{1}{3}\,\mathrm{C}\left(\frac{3P_S P_R}{N_R P_R + N_S(3P_S + N_R)}\right) \quad (5)$$

as the achievable sum rate in time slot $i$. Summing over all time slots results in (2). ∎

*Proposition 1:* For AF-IAN, the power allocation policy achieving Lemma 2 attains higher sum rates than transmitting with maximum transmit power at all nodes.

*Proof:* Using maximum transmit power in (3), the achievable sum rate with AF-IAN is

$$3\,\mathrm{C}\left(\frac{P_R P_S}{P_R P_S + P_R N_R + 3P_S N_S + N_R N_S}\right). \quad (6)$$

This rate is always less than (2) if

$$(3P_S P_R + N_R P_R + 3P_S N_S + N_S N_R)\,(P_S P_R + N_R P_R + 3P_S N_S + N_R N_S)^3 \\ - (2P_S P_R + N_R P_R + 3P_S N_S + N_R N_S)(N_R P_R + 3P_S N_S + N_S N_R) \geq 0$$

Expanding the left-hand side (LHS), we have $P_S^3 P_R^3 \left(2N_S N_R + 2N_R P_R + 3P_R P_S + 6P_S N_S\right) \geq 0$, which is easily seen to hold. ∎

*Remark 1:* Observe that time sharing is absolutely necessary to achieve the sum rate in Lemma 2. Without time sharing, the transmission scheme described above boils down to turning off one transmitter during the whole transmission. Besides being highly unfair to one user, this would also result in lower achievable rates since the remaining two transmitters are only allowed to transmit with power $P_S$. More specifically, the achievable sum rate in this case is $\text{C}\left(\frac{2P_S P_R}{N_R P_R + 2P_S N_S + N_S N_R}\right)$. It is easily shown that this rate is smaller than (2).

---

[2]Notation: $P_k'^i$ is the transmit power used by user $k$ in time slot $i$.

*Remark 2:* In each time slot of the transmission scheme described above, one user transmits with power $3P_S - p$ and the other with power $p$, where $p \in [0, 3P_S]$. However, since the sum rate does not depend on this parameter $p$, it may be chose arbitrary without affecting the sum rate. Thus, Lemma 2 included TDMA as a special case, since for $p = 0$ and $p = 3P_S$, only one user transmits per time slot.

While AF-IAN offers low relay and decoder complexity, SND [19] may be employed at the receivers to achieve higher transmission rates. In SND, the interfering message is jointly decoded with the desired message instead of treating it as additional noise.

*Lemma 3:* In the 3-user MWRC, the sum rate

$$R_\Sigma^{AF-SND} = \frac{3}{2}\,\text{C}\left(\frac{2P_S P_R}{N_R P_R + 3P_S N_S + N_S N_R}\right) \tag{7}$$

is achievable with AF relaying and SND at the receivers.

*Proof:* We start from the discrete memoryless multi-way relay channel (DM-MWRC) which is the discrete memoryless equivalent of the considered channel and model the relay using the instantaneous relaying approach from [20]. It can be seen that the resulting channel is equivalent to a 3-user interference channel (IC) with receiver message side information and feedback. Then, we apply SND to this equivalent channel which is a small variation of the proof for the 2-user IC (cf. [19, p. 135]). Finally, applying this result to Gaussian channels using the standard procedure [19, Section 3.4.1], we can show that a rate tuples $(R_1, R_2, R_3)$ is achievable in the symmetric MWRC with AF relaying and SND at the receivers if all rates are non-negative and satisfy

$$\sum_{i \in \mathcal{S}} R_i \leq \text{C}\left(\frac{P_R \sum_{i \in \mathcal{S}} P'_i}{N_R P_R + N_S \left(N_R + \sum_{j \in \mathcal{K}} P'_j\right)}\right),$$

for all $\mathcal{S} \in \{\{k\}, \{k, l(k)\} : k \in \mathcal{K}\}$. Setting $P'_k = P_S$, $k \in \mathcal{K}$, and applying the simplex algorithm [17], it is straightforward to show (7). ∎

## C. Decode-and-Forward

In DF relaying, the relay completely decodes the messages of each user and then broadcasts them back to all users. The achievable rate region is the intersection of the capacity regions of the 3-user MAC and the BC with receiver side information and partial decoding at the receivers.

*Lemma 4:* In the 3-user MWRC, the sum rate

$$R_\Sigma^{DF} = \min\left\{\frac{3}{2}\mathrm{C}\left(\frac{P_R}{N_S}\right), \mathrm{C}\left(\frac{3P_S}{N_R}\right)\right\} \qquad (8)$$

is achievable with DF relaying.

*Proof:* Evaluate the achievable rate region in [3, Proposition 2] for $L=1$, $K=3$, $P_r^1 = P_R$, $P_{1i} = P_S$ for all $i \in \mathcal{K}$, $N_r = N_R$, and $N_{1l} = N_S$ for all $l \in \mathcal{K}$. Then, using the simplex algorithm (see, e.g., [17]), we can show that

$$R_\Sigma^{DF} = \begin{cases} \frac{3}{2}D & \text{if } D < A \text{ and } C > \frac{3}{2}D, \text{ or } A < D < B \text{ and } C > \frac{3}{2}D \\ C & \text{if } D < A \text{ and } C < \frac{3}{2}D, \text{ or } A < D < B \text{ and } C < \frac{3}{2}D, \text{ or } B < D \end{cases} \qquad (9)$$

where $A = \mathrm{C}\left(\frac{P_S}{N_R}\right)$, $B = \mathrm{C}\left(\frac{2P_S}{N_R}\right)$, $C = \mathrm{C}\left(\frac{3P_S}{N_R}\right)$, and $D = \mathrm{C}\left(\frac{P_R}{N_S}\right)$.

We observe that $B < D \Leftrightarrow \frac{2P_S}{N_R} < \frac{P_R}{N_S}$. Thus, it holds that $\left(1+3\frac{P_S}{N_R}\right)^2 < \left(1+\frac{3}{2}\frac{P_R}{N_S}\right)^2 < \left(1+\frac{P_R}{N_S}\right)^3$ which is equivalent to $C < \frac{3}{2}D$. Hence, with $B < D \Rightarrow C < \frac{3}{2}D$ it can be seen from (9) that $R_\Sigma^{DF} = \min\left\{\frac{3}{2}D, C\right\}$. ∎

*Remark 3:* The result from [3] implements a full message exchange. However, from the proof of Lemma 1 it can be seen that in the symmetric case the relaxed decoding requirements due to the partial message exchange considered here can not result in higher rates for DF.

### D. Noisy Network Coding

NNC [21] generalizes the compress-and-forward (CF) coding scheme for the relay channel [22] and network coding for graphical networks [23] to discrete memoryless networks (DMNs).

*Lemma 5:* In the 3-user MWRC, the sum rate

$$R_\Sigma^{NNC} = \frac{3}{2}\mathrm{C}\left(\frac{2P_S P_R}{N_R P_R + 2P_S N_S + N_S N_R}\right) \qquad (10)$$

is achievable with NNC and SND.

*Proof:* Use [21, Theorem 2] and identify $\mathcal{D}_0 = \emptyset$ and $\mathcal{D}_k = \{q(k)\}$ for $k \in \mathcal{K}$. Assume $\hat{Y}_i = Y_i + \hat{Z}_i$ with $\hat{Z}_i \sim \mathcal{CN}(0, Q_i)$ for $i \in \mathcal{K}_R$, and $Q = \emptyset$, i.e., no time-sharing is used. Then, the achievable rate region is

$$R_k < \mathrm{C}\left(\frac{P_S}{N_R + Q_0}\right), \qquad \sum_{i \in \mathcal{K}\setminus\{k\}} R_i < \min\left\{\mathrm{C}\left(\frac{2P_S}{N_R + Q_0}\right), \mathrm{C}\left(\frac{P_R}{N_S}\right) - \mathrm{C}\left(\frac{N_R}{Q_0}\right)\right\},$$

for each $k \in \mathcal{K}$.

Using the simplex algorithm [17], it is obtained that the sum rate is $R_\Sigma = \frac{3}{2}\text{C}\left(\frac{2P_S}{N_R+Q_0}\right)$ if $Q_0 > \frac{N_S(N_R+2P_S)}{P_R}$, and $R_\Sigma = \frac{3}{2}\text{C}\left(\frac{P_R}{N_S}\right) - \text{C}\left(\frac{N_R}{Q_0}\right)$, otherwise. Due to monotonicity in $Q_0$, the optimal choice for $Q_0$ is $\frac{N_S(N_R+2P_S)}{P_R}$. Evaluating $R_\Sigma$ at this point results in (10). ∎

### E. Comparisons and Insights

In this subsection, we use the results from above to state some analytical results on the sum capacity and compare achievable sum rates of the considered relaying schemes.

*Theorem 1:* If $\frac{P_R}{N_S} \leq \left(1+3\frac{P_S}{N_R}\right)^{2/3} - 1$, the sum capacity of the symmetric 3-user MWRC is $C_\Sigma = \frac{3}{2}\text{C}\left(\frac{P_R}{N_S}\right)$ and is achieved using the DF strategy from Lemma 4.

*Proof:* Equation (8) is equal to (1) if in both equations the first term of the minimum is active. Since $\text{C}\left(\frac{3P_S}{N_R}\right) \leq 3\,\text{C}\left(\frac{P_S}{N_R}\right)$, this is the case as long as $\frac{3}{2}\text{C}\left(\frac{P_R}{N_S}\right) \leq \text{C}\left(\frac{3P_S}{N_R}\right)$. This condition is equivalent to the condition in Theorem 1. ∎

*Remark 4:* For the completely symmetric channel where $P = P_S = P_R$ and $N = N_S = N_R$, the condition in Theorem 1 is equivalent to $\frac{P}{N} \leq 3 + 2\sqrt{3} \approx 8.1\,\text{dB}$.

*Theorem 2:* The sum capacity of the completely symmetric 3-user MWRC is achieved within $0.877\,\text{bit}$ by NNC and within $1.5\,\text{bit}$ by AF-SND.

*Proof:* Let $S = \frac{P_R}{N_S} = \frac{P_S}{N_R}$. The achievable sum rate for NNC and AF-SND is $\text{C}\left(\frac{2S^2}{1+(1+a)S}\right)$ with $a=2$ for NNC and $a=3$ for AF. Its difference to the outer bound is

$$\Delta = \frac{3}{2}\text{C}(S) - \frac{3}{2}\text{C}\left(\frac{2S^2}{1+(1+a)S}\right) = \frac{3}{2}\log\left(\frac{1+(2+a)S+(1+a)S^2}{1+(1+a)S+2S^2}\right).$$

From the first derivative of $\Delta$ it can be seen that this is an increasing function in $S$ for the relevant choices of $a$. Thus, its maximum is attained at $S \to \infty$ and has value $\lim_{S \to \infty} \Delta = \frac{3}{2}\log\left(\frac{1+a}{2}\right)$. ∎

*Corollary 1:* The sum degrees of freedom (DoF) of the 3-user MWRC with circular message exchange is 1.5.

*Proof:* The sum DoF $d_\Sigma$ is defined as $\lim_{P \to \infty} \frac{C_\Sigma(P)}{\log(P)}$ where $C_\Sigma(P)$ is the sum capacity of the completely symmetric MWRC. From the proof of Theorem 2, for the outer and inner bounds we respectively have

$$d_\Sigma^{out} \leq \lim_{S \to \infty} \frac{3}{2}\frac{\log(1+S)}{\log(S)} = \frac{3}{2} \qquad d_\Sigma^{in} \geq \lim_{S \to \infty} \frac{3}{2}\frac{\log\left(1+\frac{2S^2}{1+(1+a)S}\right)}{\log(S)} = \frac{3}{2}.$$

Since both bounds coincide, $d_\Sigma = \frac{3}{2}$. ∎

*Remark 5:* It is easily shown that DF and AF-IAN achieve a sum DoF of 1.

*Proposition 2:* It holds that $R_\Sigma^{NNC} \geq R_\Sigma^{AF-SND} \geq R_\Sigma^{AF-IAN}$.

*Proof:* The first inequality follows directly from comparing (7) and (10). For the second, note that $R_\Sigma^{AF-SND} \geq R_\Sigma^{AF-IAN}$ is equivalent to

$$\left(1 + \frac{2P_S P_R}{N_R P_R + 3P_S N_S + N_S N_R}\right)^3 \geq \left(1 + \frac{3P_S P_R}{N_R P_R + 3P_S N_S + N_S N_R}\right)^2.$$

Since $f(x) = (1+2x)^3 - (1+3x)^2$ is monotonically increasing and $f(0) = 0$, we have $f(x) \geq 0$ for $x \geq 0$ and, hence, the above inequality holds. ∎

*Remark 6:* DF does not fit in this order since it outperforms NNC only in certain SNR regimes. Even for the completely symmetric case, the intersection point between DF and NNC has to be determined numerically. In this case it is approximately 14.27 dB.

## IV. RESOURCE ALLOCATION FOR ENERGY EFFICIENCY MAXIMIZATION

From a physical standpoint, the efficiency in the use of a given resource is the benefit-cost ratio, i.e. the benefit obtained from using the resource, divided by the cost associated to the use of the resource. Applying this general concept to the use of energy to transmit data over a communication link, leads to defining the EE as the amount of data that can be reliably transmitted in a given time interval $T$, divided by the resulting energy consumption over the same time interval.

In this section, we will deal with the problem of allocating the transmit powers at the users and relay for EE maximization. Both cooperative and competitive resource allocation will be considered.

### A. Cooperative resource allocation

From a system-wide perspective, the benefit and cost related to the use of energy over a time interval $T$ are identified as the amount of data that can be reliably transmitted in the whole network during $T$, and the total energy consumption in the network. This leads to defining the system GEE as [8], [24]

$$\text{GEE} = \frac{TBR_\Sigma}{T(\phi P_S + \psi P_R + P_c)} = \frac{BR_\Sigma}{\phi P_S + \psi P_R + P_c}. \tag{11}$$

In (11), $B$ is the communication bandwidth and $R_\Sigma$ is the achievable sum rate in bit/s/Hz, which depends on the particular relay protocol and receive scheme. Therefore, the numerator $TBR_\Sigma$

TABLE I
PARAMETERS FOR GLOBAL ENERGY EFFICIENCY MAXIMIZATION.

(a) GEE$_1$

| Scheme | $a_1$ | $\alpha_1$ | $a_2$ | $\alpha_2$ |
|---|---|---|---|---|
| Outer bound | $\frac{3}{2}$ | 1 | 3 | 1 |
| DF | $\frac{3}{2}$ | 1 | 1 | 3 |

(b) GEE$_2$

| Scheme | $\alpha$ | $a$ | $b$ | $c$ |
|---|---|---|---|---|
| AF-SND | $\frac{3}{2}$ | $\frac{3}{2}N_S$ | $\frac{1}{2}N_R$ | $\frac{1}{2}N_S N_R$ |
| AF-IAN | 1 | $N_S$ | $\frac{1}{3}N_R$ | $\frac{1}{3}N_S N_R$ |
| NNC | $\frac{3}{2}$ | $N_S$ | $\frac{1}{2}N_R$ | $\frac{1}{2}N_S N_R$ |

is the total amount of data that can be reliably transmitted during $T$. On the other hand, the associated total energy consumption is given by the sum of the energies consumed for signal transmission at the users and relay plus the circuit power that is dissipated in all terminals to operate the devices. The first component is expressed as $\phi P_S + \psi P_R$, with $\phi$ and $\psi$ being the inefficiencies of the users and relay transmit amplifiers,[3] while the second component is modeled by the constant term $P_c$.

In the cooperative scenario, the users and the relay cooperate and jointly allocate $P_S$ and $P_R$ to maximize the common performance metric (11). Given the achievable sum rate expressions from Section III, (11) takes two different functional forms. For the outer bound and for the DF case we have GEE$_1$, and for the AF and NNC cases we have GEE$_2$ as defined next:

$$\text{GEE}_1 = \frac{\min\left\{a_1 \, \text{C}\left(\alpha_1 \frac{P_R}{N_S}\right), a_2 \, \text{C}\left(\alpha_2 \frac{P_S}{N_R}\right)\right\}}{\phi P_S + \psi P_R + P_c}, \qquad \text{GEE}_2 = \frac{\alpha \, \text{C}\left(\frac{P_S P_R}{a P_S + b P_R + c}\right)}{\phi P_S + \psi P_R + P_c}, \qquad (12)$$

with $a_1$, $a_2$, $\alpha_1$, $\alpha_2$, $\alpha$, $a$, $b$, and $c$ non-negative parameters.

Being fractional functions, the considered GEEs are in general non-concave functions and conventional convex programming tools can not be used. Instead, fractional programming provides a framework to optimize fractional functions. In particular, we will exploit the following result from fractional programming theory.

*Proposition 3:* Let $\mathcal{S} \in \mathbb{R}^n$, $f, g : \mathcal{S} \to \mathbb{R}$, with $f(\boldsymbol{x}) \geq 0$ and $g(\boldsymbol{x}) > 0$. Solving the problem $\max_{\boldsymbol{x} \in \mathcal{S}} \frac{f(\boldsymbol{x})}{g(\boldsymbol{x})}$ is equivalent to finding the unique zero of the function $F(\lambda) = \max_{\boldsymbol{x} \in \mathcal{S}} (f(\boldsymbol{x}) - \lambda g(\boldsymbol{x}))$.

*Proof:* See [25], [26]. ∎

Proposition 3 provides a way to maximize a fractional function by finding the zero of the auxiliary

---

[3]Specifically, $\psi \geq 1$ is the inefficiency of the relay amplifier, while $\phi \geq 3$, accounts for the inefficiency of the amplifiers of the three users, assumed equal for notational convenience

function $F(\lambda)$. This can be accomplished by means of Dinkelbach's algorithm [26], which is reported in Algorithm 1.

**Algorithm 1** Dinkelbach's Algorithm
---
1: Initialize $\lambda_0$, such that $F(\lambda_0) \geq 0$. Set $n = 0$.
2: **while** $F(\lambda_n) > \epsilon$ **do**
3:
$$\boldsymbol{x}_n^\star \leftarrow \arg\max_{\boldsymbol{x} \in \mathcal{S}} f(\boldsymbol{x}) - \lambda_n g(\boldsymbol{x}) \tag{13}$$
4: $\quad \lambda_{n+1} \longleftarrow = \frac{f(\boldsymbol{x}_n^\star)}{g(\boldsymbol{x}_n^\star)}$
5: $\quad n \leftarrow n + 1$
6: **end while**
7: Output $(\boldsymbol{x}_n^\star, \lambda_n)$
---

Dinkelbach's algorithm exhibits super-linear convergence and only requires the solution of a sequence of convex problems, provided $f(\boldsymbol{x})$ and $g(\boldsymbol{x})$ are concave and convex, respectively, and that $\mathcal{S}$ is a convex set. Indeed, in this case the subproblem to be solved in each iteration to find $\boldsymbol{x}_n^*$ is convex. Moreover, we stress that Dinkelbach's algorithm converges to the global solution of the associated fractional problem also when $f(\boldsymbol{x})$ is not concave and/or $g(\boldsymbol{x})$ is not convex, even though in this case a non-convex problem must be globally solved in each iteration.

*1) Maximization of* $\text{GEE}_1$: The optimization problem is formulated as

$$\begin{cases} \max_{P_S, P_R} \dfrac{\min\left\{a_1 \, \text{C}\left(\alpha_1 \frac{P_R}{N_S}\right), a_2 \, \text{C}\left(\alpha_2 \frac{P_S}{N_R}\right)\right\}}{\phi P_S + \psi P_R + P_c} \\ \text{s.t.} \quad P_S \in [0; P_S^{max}], \ P_R \in [0; P_R^{max}]. \end{cases} \tag{14}$$

Problem (14) is an instance of a non-concave and non-smooth fractional problem. However, it can be efficiently solved by means of Dinkelbach's algorithm because the minimum of concave functions is concave. Therefore, the objective of (14) has a concave numerator and an affine denominator. Moreover, it is possible to reformulate (14) into a smooth problem by introducing the auxiliary variable $t$ as follows.

$$\begin{cases} \max_{t, P_S, P_R} \dfrac{t}{\phi P_S + \psi P_R + P_c} \\ \text{s.t.} \quad P_S \in [0; P_S^{max}], \ P_R \in [0; P_R^{max}], \ a_1 \, \text{C}\left(\alpha_1 \dfrac{P_R}{N_S}\right) - t \geq 0, \ a_2 \, \text{C}\left(\alpha_2 \dfrac{P_S}{N_R}\right) - t \geq 0. \end{cases} \tag{15}$$

In (15) the numerator and denominator of the objective are both linear, while the constraints are convex. Then, (15) can be solved by means of Dinkelbach's algorithm with an affordable complexity.

*2) Maximization of* $\text{GEE}_2$: In this case, the optimization problem is formulated as

$$\begin{cases} \max_{P_S, P_R} \dfrac{\alpha \, \text{C}\left(\dfrac{P_S P_R}{a P_S + b P_R + c}\right)}{\phi P_S + \psi P_R + P_c} \\ \text{s.t.} \quad P_S \in [0; P_S^{max}], \quad P_R \in [0; P_R^{max}]. \end{cases} \quad (16)$$

Problem (16) is more challenging than Problem (14) because the numerator of the objective function is not jointly concave in the optimization variables. Therefore, directly applying Dinkelbach's algorithm requires solving a sequence of non-convex problem, which is computationally demanding. In general, no optimization tool is available to maximize a fractional function with non-concave numerator with limited complexity [25]. As a consequence, finding the global solution of (16) with limited complexity appears difficult. In the following, we propose two approaches to solve (16). The first approach will be based on the alternating maximization algorithm [27] and will find a local optimum of (16) with a limited computational complexity. Instead, the second approach will leverage the theory of monotonic optimization, determining the global solution of (16) with an exponential complexity. Interestingly, numerical evidence to be provided in Section VI will show that the two algorithms achieve virtually equal performance.

*a) Alternating maximization approach:* This approach is based on the observation that the numerator of the objective, although not being jointly concave in $P_S$ and $P_R$, is separately concave in the two variables. To be more specific, the numerator of the objective is concave in $P_S$ for fixed $P_R$ and vice versa. Therefore, if we fix either $P_S$ or $P_R$, we can solve (16) with respect to the other variable by applying Dinkelbach's algorithm and solving a sequence of convex problems. The formal algorithm is reported next and labeled Algorithm 2.

---

**Algorithm 2** Alternating maximization for Problem (16)

```
Initialize P_R^(0) ∈ [0,P_R^max]. Set a tolerance ε. Set n = 0.
```
**while** $\left|\text{EE}_2^{(n)} - \text{EE}_2^{(n-1)}\right| \geq \epsilon$ **do**
```
    P_S^(n+1) ← Solve Problem (16) with respect to P_S for fixed P_R^(n).
    P_R^(n+1) ← solve Problem (16) with respect to P_R for fixed P_S^(n+1).
    n ← n + 1
```
**end while**
```
Output (P_S^(n), P_R^(n)).
```

---

The following proposition holds.

*Proposition 4:* Algorithm 2 converges to a stationary point of Problem (16).

*Proof:* After each iteration of Algorithm 2 the objective is not decreased. Hence, convergence follows since the objective is upper bounded. Convergence to a stationary point follows by leveraging [27, Proposition 2.7.1], which states that alternating maximization converges to a stationary point if: 1) the feasible set is the Cartesian product of closed and convex sets; 2) the objective is continuously differentiable on the feasible set; 3) the solution to each subproblem is unique. In our case, 1) and 2) are apparent. As for Assumption 3), it follows from the properties of fractional functions. In particular, when either $P_S$ or $P_R$ is fixed, the numerator of the objective is strictly concave in the other variable, as can be seen by computing the second derivative (see also [12]). Then, since the ratio between a strictly concave and an affine function is known to be strictly pseudo-concave, each subproblem in Algorithm 2 has a strictly pseudo-concave objective. Finally, the thesis follows because a strictly pseudo-concave function admits a unique maximizer. ∎

*b) Monotonic optimization approach:* As mentioned, when the concave/convex structure of the objective does not hold, Dinkelbach's algorithm still converges to the global solution of the original fractional problem, provided one can globally solve the non-convex Problem (13) in each iteration. For the case at hand, Problem (13) takes the form

$$\begin{cases} \max_{P_S, P_R} \alpha \, \mathrm{C}\left(\dfrac{P_S P_R}{a P_S + b P_R + c}\right) - \lambda\left(\phi P_S + \psi P_R + P_c\right) \\ \text{s.t.} \quad P_S \in [0; P_S^{max}], \quad P_R \in [0; P_R^{max}]. \end{cases} \quad (17)$$

Although being non-convex, Problem (17) can be globally solved by means of monotonic optimization theory. For a detailed review on monotonic optimization we refer to [28], [29]. Here we only remark that monotonic optimization provides a framework to globally maximize monotone functions over normal sets, and recall the following facts.

*Definition 1:* Let $\boldsymbol{x}_{min}, \boldsymbol{x}_{max} \in \mathbb{R}_+^n$ with[4] $\boldsymbol{x}_{min} \leq \boldsymbol{x}_{max}$. Then, $[\boldsymbol{x}_{min}, \boldsymbol{x}_{max}] = \{\boldsymbol{x} \in \mathbb{R}_+^n : \boldsymbol{x}_{min} \leq \boldsymbol{x} \leq \boldsymbol{x}_{max}\}$ is a hyper-rectangle in $\mathbb{R}_+^n$ and the set $\mathcal{S} \subset \mathbb{R}_+^n$ is normal if $[\boldsymbol{0}, \boldsymbol{x}] \in \mathcal{S}$, $\forall \, \boldsymbol{x} \in \mathcal{S}$.

*Proposition 5:* The set $\mathcal{S} = \{\boldsymbol{x} \in \mathbb{R}_+^n : g(\boldsymbol{x}) \leq 0\}$ is normal and closed if $g : \mathbb{R}_+^n \to \mathbb{R}$ is lower semi-continuous and increasing.

---

[4]Inequalities between vectors are component-wise.

The feasible set of (17) is normal, being the cartesian product of two closed convex intervals in $\mathbb{R}$, but the objective function is not monotone, as it is the difference of two increasing functions. However, the following proposition shows how to reformulate (17) into a monotonic problem.

*Proposition 6:* Problem (17) can be equivalently reformulated as

$$\begin{cases} \max_{P_S, P_R, t} \alpha \, \mathrm{C}\left(\dfrac{P_S P_R}{aP_S + bP_R + c}\right) + t \\ \text{s.t. } P_S \in [0; P_S^{max}], \ P_R \in [0; P_R^{max}], \ t \geq 0, \ t + \lambda(\phi P_S + \psi P_R) \leq \lambda(\phi P_S^{max} + \psi P_R^{max}). \end{cases} \quad (18)$$

Problem (18) is a monotonic problem in canonical form.

*Proof:* Introducing the auxiliary variable $t$ and using the substitution $\lambda \left( \phi P_S + \psi P_R + P_c \right) + t = \lambda \left( \phi P_S^{max} + \psi P_R^{max} + P_c \right)$, yields the objective of (18), up to the inessential constant $-\lambda \left( \phi P_S^{max} + \psi P_R^{max} + P_c \right)$. Next, since $\lambda \left( \phi P_S + \psi P_R + P_c \right)$ is non-negative and increasing in both $P_S$ and $P_R$, the considered substitution is equivalent to $t \geq 0$, $t + \lambda \left( \phi P_S + \psi P_R \right) \leq \lambda \left( \phi P_S^{max} + \psi P_R^{max} \right)$, and we obtain Problem (18). Finally, we have to show that (18) is a monotonic problem. The monotonicity of the objective follows because the function $\mathrm{C}(\cdot)$ is increasing and its argument $\frac{P_S P_R}{aP_S + bP_R + c}$ can be shown to be increasing in both $P_S$ and $P_R$ by direct computation of first-order derivatives. Moreover, since the constraint function $t + \lambda \left( \phi P_S + \psi P_R \right)$ is increasing, the feasible set is contained in the hyper-rectangle $[0; P_S^{max}] \times [0; P_R^{max}] \times [0; \lambda(\phi P_S^{max} + \psi P_R^{max})]$. Moreover, the feasible set is also normal, by virtue of Proposition 5. ∎

As a consequence of Proposition 6, we can apply Dinkelbach's algorithm to globally solve (16), where in each iteration the solution to the subproblem (17) is found by solving the equivalent problem (18) by standard monotonic optimization algorithms, such as the polyblock algorithm [29]. The drawback of this approach is that the complexity of monotone programming is in general exponential. However, we stress two points. First, in our scenario we only have three variables, $P_S$, $P_R$, and the auxiliary variable $t$, which makes the use of monotonic optimization viable. Second, the proposed method provides a benchmark for the proposed low-complexity algorithm based on alternating optimization. As already mentioned, our numerical results will show that the two methods enjoy very similar performance.

## B. Competitive resource allocation

After analyzing the cooperative case, we turn our attention to the competitive scenario in which a distributed power allocation is performed. This scenario is particularly relevant for distributed networks in which no central coordination is employed and power allocation needs to be performed in a distributed fashion. More in detail, in this section we consider the case in which $P_S$ and $P_R$ are not jointly allocated to maximize the GEE, but rather are separately allocated in a competitive fashion for individual EE maximization. We observe that in the considered symmetric scenario where the three users use a common power level $P_S$, the competition occurs between the relay, which allocates $P_R$, and the collective of the three transmitters, which together allocate $P_S$. This competitive power control problem can be formulated as a non-cooperative game in normal form $\mathcal{G}$, whose players are the relay $\mathcal{R}$, with strategy set $[0; P_R^{max}]$, and the three the sources together $\{\mathcal{S}_k\}_{k \in \mathcal{K}}$, with strategy set $[0; P_S^{max}]$. The players' utility functions are the individual EEs, defined as

$$u_S = \frac{R_\Sigma}{P_S + P_{c,S}}, \qquad u_R = \frac{R_\Sigma}{P_R + P_{c,R}}, \tag{19}$$

with $P_{c,S}$ and $P_{c,R}$ the energy consumptions of the three sources and of the relay, respectively.[5] We observe that both the EEs in (19) have the achievable sum rate at the numerator. This is motivated by the following observation: since the users allocate the same power level $P_S$, they should tune $P_S$ so as to optimize their collective benefit, i.e. the achievable sum rate $R_\Sigma$. On the other hand, the relay benefit is also represented by the achievable sum rate, since the relay typically does not favor any particular user. Moreover, in many circumstances, the relay is deployed by a third-party operator, which loans the relay to the users, and whose revenue is proportional to the total amount of conveyed data.

The BRD of $\mathcal{G}$ is expressed as the following two coupled problems $\max_{P_i} \frac{R_\Sigma(P_i, P_{-i})}{P_i + P_{c,i}}$, with $P_i \in [0; P_i^{max}]$, $i \in \{S, R\}$, and $P_{-i}$ denoting the strategy of the player other than $i$. The optimization problem for $i = S$ is the best response (BR) of the sources to the relay strategy, whereas the problem for $i = R$ is the BR of the relay to the sources strategy. Depending on the relay operation, $R_\Sigma$ can take one of the two following general forms

$$R_{\Sigma,1} = \min\left\{a_1 \, \text{C}\left(\alpha_1 \frac{P_R}{N_S}\right), a_2 \, \text{C}\left(\alpha_2 \frac{P_S}{N_R}\right)\right\}, \qquad R_{\Sigma,2} = \alpha \, \text{C}\left(\frac{P_S P_R}{a P_S + b P_R + c}\right), \tag{20}$$

with $a_1$, $a_2$, $\alpha_1$, $\alpha_2$, $\alpha$, $a$, $b$, and $c$ non-negative parameters as defined in Table I. In the following, we will denote by $\mathcal{G}_1$ and $\mathcal{G}_2$ the game with $R_{\Sigma,1}$ and $R_{\Sigma,2}$, respectively. Moreover, for ease of notation, let us define $\gamma_1 = \frac{\alpha_1 P_R}{N_S}$, $\gamma_2 = \frac{\alpha_2 P_S}{N_R}$, and $\gamma = \frac{P_S P_R}{a P_S + b P_R + c}$. We start our analysis with the following result.

---

[5]We have not considered any scaling coefficients in front of the transmit powers. This entails no loss of generality since any scaling coefficient can be included in the circuit power terms without affecting the maximizers of the utilities.

*Proposition 7:* $\mathcal{G}_1$ and $\mathcal{G}_2$ always admit an NE. Moreover, denote by $\mathrm{BR}_S(P_R)$ and $\mathrm{BR}_R(P_S)$ the BRs of the sources and of the relay, respectively. In case of $\mathcal{G}_1$ it holds

$$\mathrm{BR}_S(P_R) = \min(P_S^{max}, \bar{P}_S^{max}, \bar{P}_S), \qquad \mathrm{BR}_R(P_S) = \min(P_R^{max}, \bar{P}_R^{max}, \bar{P}_R), \qquad (21)$$

with $\bar{P}_S^{max} = \frac{N_R}{\alpha_2}[(1+\frac{\alpha_1 P_R}{N_S})^{a_1/a_2} - 1]$, $\bar{P}_R^{max} = \frac{N_S}{\alpha_1}[(1+\frac{\alpha_2 P_S}{N_R})^{a_2/a_1} - 1]$, $\bar{P}_S$ the unique solution of the equation $\frac{N_R\,\mathrm{C}(\gamma_2)}{\alpha_2\,\mathrm{C}'(\gamma_2)} - P_S = P_{c,S}$, and $\bar{P}_R$ the unique solution of the equation $\frac{N_S\,\mathrm{C}(\gamma_1)}{\alpha_1\,\mathrm{C}'(\gamma_1)} - P_R = P_{c,R}$. In case of $\mathcal{G}_2$ it holds

$$\mathrm{BR}_S(P_R) = \min(P_S^{max}, \bar{P}_S), \qquad \mathrm{BR}_R(P_S) = \min(P_R^{max}, \bar{P}_R), \qquad (22)$$

with $\bar{P}_S$ the unique solution of the equation $\frac{\mathrm{C}(\gamma)}{\mathrm{C}'(\gamma)d\gamma/dP_S} - P_S = P_{c,S}$ and $\bar{P}_R$ the unique solution of the equation $\frac{\mathrm{C}(\gamma)}{\mathrm{C}'(\gamma)d\gamma/dP_R} - P_R = P_{c,R}$.

*Proof:* The existence follows by observing that for both $\mathcal{G}_1$ and $\mathcal{G}_2$ the strategy sets are closed and compact [30], the utility functions are continuous in $P_S$ and $P_R$, and that $u_S$ is pseudo-concave in $P_S$, while $u_R$ is pseudo-concave in $P_R$.

Next, let us focus on $\mathcal{G}_1$ and on the derivation of $\mathrm{BR}_S(P_R)$. For any fixed $P_R$, the $P_S$ where $a_1\,\mathrm{C}(\gamma_1)$ intersects $a_2\,\mathrm{C}(\gamma_2)$ is $\bar{P}_S^{max}$. A key observation is that the optimal $P_S$ will never be larger than $\bar{P}_S^{max}$, because for $P_S \geq \bar{P}_S^{max}$ the numerator of $u_S$ keeps constant, while the denominator increases. On the other hand, in the range $[0; \bar{P}_S^{max}]$, we have $u_S = a_2\frac{\mathrm{C}(\gamma_2)}{P_S+P_{c,S}}$, which is increasing in $[0; \bar{P}_S]$, with $\bar{P}_S$ its unique maximizer. This proves $\mathrm{BR}_S(P_R)$ in (21) and by a similar reasoning we can prove $\mathrm{BR}_R(P_S)$.

Finally, let us focus on $\mathcal{G}_2$ and on the derivation of $\mathrm{BR}_S(P_R)$. In this case, the utility functions are differentiable, which allows to obtain $\bar{P}_S$ from the first-order optimality condition $\frac{\mathrm{C}(\gamma)}{\mathrm{C}'(\gamma)d\gamma/dP_S} - P_S = P_{c,S}$. Since $u_S$ is also pseudo-concave in $P_S$, first-order conditions are also sufficient and $\bar{P}_S$ is the global maximizer of $u_S$ for any fixed $P_R$. Accounting for the maximum power constraint we obtain $\mathrm{BR}_S(P_R)$ in (22), while $\mathrm{BR}_R(P_S)$ follows by a similar argument. ∎

After establishing the existence of at least one NE, a natural question to ask is whether a unique NE exists and whether the BRD converges to an NE. Unlike previous related results on non-cooperative power control in two-hop systems [31], [12], which show the uniqueness of the NE for the scenario in which the relay is not one player of the game, for the case at hand uniqueness does not hold. To see this, first observe that the strategy profile $(P_S = 0, P_R = 0)$ is an NE. Indeed, if either $P_S$ or $P_R$ is set to zero, then the utility functions are both identically zero and the other player has no incentive to use a non-zero transmit power. This circumstance is a direct consequence of the fact that the relay has been included as a player of the game, a scenario which was not considered in previous works where the relay was assumed to have no EE concerns. Next, it is easy to find numerical examples in which also non-trivial NE exist, thus proving that more than one NE exists in general. However, while the uniqueness of the NE is lost if the relay takes part in the game, the convergence of the BRD still holds. To show this, the following lemma is required, which proves the monotonicity of the BR functions of $\mathcal{G}_1$ and $\mathcal{G}_2$.

*Lemma 6:* For $\mathcal{G}_1$ and $\mathcal{G}_2$, $\mathrm{BR}_S(P_R)$ is increasing in $P_R$ and $\mathrm{BR}_R(P_S)$ is increasing in $P_S$.

*Proof:* Let us consider $\mathcal{G}_1$ first. We will show the monotonicity of $\mathrm{BR}_S(P_R)$. The monotonicity of $\mathrm{BR}_R(P_S)$ can be obtained by a similar argument. From Proposition 7 we know that $\mathrm{BR}_S(P_R) = \min(P_S^{max}, \bar{P}_S^{max}, \bar{P}_S)$. The only argument of the minimum function that depends on $P_R$ is $\bar{P}_S^{max}$, since the equation whose solution defines $\bar{P}_S$ does not actually depend on $P_R$. Moreover, $\bar{P}_S^{max}$, whose expression has been provided in Proposition 7, is an increasing function of $P_R$. Consequently, recalling that the $\min$ function is increasing, we conclude that $\mathrm{BR}_S(P_R)$ is increasing in $P_R$.

Let us now consider $\mathcal{G}_2$. The BR of the sources to the relay strategy is $\mathrm{BR}_S(P_R) = min(P_S^{max}, \bar{P}_S)$, with $\bar{P}_S$ obtained by solving

$$\frac{\mathrm{C}(\gamma)}{\mathrm{C}'(\gamma) d\gamma/dP_S} - P_S = P_{c,S} \tag{23}$$

Let us define the LHS of (23) as the function $g(P_S, P_R)$. The first step of the proof is to show that $g$ is increasing in $P_S$.

$$\frac{\partial g}{\partial P_S} = -\frac{\mathrm{C}(\gamma)\left[\mathrm{C}''(\gamma)(\gamma'(P_S))^2 + \gamma''(P_S)\mathrm{C}'(\gamma)\right]}{\left(\mathrm{C}'(\gamma)\gamma'(P_S)\right)^2} \geq 0 \,, \tag{24}$$

where we have exploited the fact that $\mathrm{C}(\gamma)$ and $\gamma(P_S)$ are positive, increasing, and concave. Next, we will show that $g$ is decreasing in $P_R$.

$$\frac{\partial g}{\partial P_R} \leq 0 \iff \left(\mathrm{C}'(\gamma)\right)^2 - \mathrm{C}(\gamma)\mathrm{C}''(\gamma) \leq \mathrm{C}(\gamma)\mathrm{C}'(\gamma)\frac{\partial^2 \gamma}{\partial P_S P_R}\frac{1}{\frac{\partial \gamma}{\partial P_R}\frac{\partial \gamma}{\partial P_S}} \tag{25}$$

Elaborating, the last two terms of the RHS can be expressed as

$$\frac{\partial^2 \gamma}{\partial P_S P_R} \frac{1}{\frac{\partial \gamma}{\partial P_R} \frac{\partial \gamma}{\partial P_S}} = \frac{2abP_SP_R + acP_S + cbP_R + c^2}{(bP_R+c)(aP_S+c)} \left( \frac{aP_S + bP_R + c}{P_S P_R} \right)$$

$$= \left( 1 + \left( 1 + \frac{c}{ab} \frac{aP_S + bP_0 + c}{P_S P_R} \right)^{-1} \right) \frac{1}{\gamma} = \frac{1}{\gamma} + \frac{1}{\gamma + d},$$

with $d = c/ab$. Thus, (25) becomes $\left( C'(\gamma) \right)^2 - C(\gamma) C''(\gamma) \leq C(\gamma) C'(\gamma) \left( \frac{1}{\gamma} + \frac{1}{\gamma+d} \right)$, and plugging the expressions of $C(\gamma)$ and its derivatives, we have

$$\gamma \leq \log(1+\gamma) + \frac{\gamma(1+\gamma)}{\gamma + d} \log(1+\gamma) \tag{26}$$

It can be seen that (26) holds if $d \leq 1$. In particular, for $d = 1$ we have $\gamma \leq (\gamma+1)\log(1+\gamma)$. Both functions start from zero, but the RHS has a larger derivative for all $\gamma \geq 0$. Then, (26) holds also for $d < 1$ since the RHS is decreasing in $d$. For both AF-IAN and NNC we have $d = 1$, while for AF-SND we have $d = 2/3$. Hence, for all of the considered schemes, (26) is true and if $P_R$ increases, the LHS of (23) decreases. As a consequence, in order to reach the constant level $P_{c,S}$ at the RHS, $\bar{P}_S$ must increase, since we have also shown that the LHS of (23) is increasing in $P_S$. Finally, the monotonicity of the BR follows from the fact that the min function is increasing. ■

*Proposition 8:* For any set of system parameters and initialization point, the BRDs of both $\mathcal{G}_1$ and $\mathcal{G}_2$ are guaranteed to converge.[6]

*Proof:* Denote by $P_S^{(0)}$ the initial value of $P_S$. Then, $P_R^{(0)} = \text{BR}_R(P_S^{(0)})$ and $P_S^{(1)} = \text{BR}_S(P_R^{(0)})$. Let us consider three cases.

If $P_S^{(1)} = P_S^{(0)}$, we have reached convergence.

If $P_S^{(1)} > P_S^{(0)}$, then we also have $P_R^{(1)} = \text{BR}_R(P_S^{(1)}) \geq P_R^{(0)}$ by virtue of Lemma 6. But then, at the second iteration we also have $P_S^{(2)} = \text{BR}_S(P_R^{(1)}) \geq P_S^{(1)}$ and $P_R^{(2)} = \text{BR}_R(P_S^{(2)}) \geq P_R^{(1)}$. Similarly, at the $n$-th iteration we have $P_S^{(n)} \geq P_S^{(n-1)}$ and $P_R^{(n)} \geq P_R^{(n-1)}$. Since the sources and relay BRs are upper bounded by $P_S^{max}$ and $P_R^{max}$, respectively, the iteration must converge.

If $P_S^{(1)} < P_S^{(0)}$, by a similar argument it follows that after each iteration $P_S$ and $P_R$ are not increased, i.e. $P_S^{(n)} \leq P_S^{(n-1)}$ and $P_R^{(n)} \leq P_R^{(n-1)}$. Thus, convergence must eventually occur since the BRs are non negative. ■

---

[6]Convergence is meant in the value of the utilities.

Resource Allocation for Energy-Efficient 3-Way Relay Channels 22

Based on Proposition 8, the BRD-based, competitive, power control algorithm in which each player iteratively maximizes his own utility function is guaranteed to converge. The formal procedure is reported next and labeled Algorithm 3.

---

**Algorithm 3** Competitive power control algorithm
---

```
Initialize P_R^(0) ∈ [0,P_R^max]. Set a tolerance ϵ.
```
**while** $\max_{i=S,R} \left\{ |\text{EE}_i^{(n)} - \text{EE}_i^{(n-1)}| \right\} \geq \epsilon$ **do**
```
    P_S^(n+1) ← min(P_S^max,P̄_S) for given P_R^(n), with P̄_S defined in Proposition 7
    P_R^(n+1) ← min(P_R^max,P̄_R) for given P_S^(n+1), with P̄_R defined in Proposition 7
```
**end while**
```
Output (P_S^(n),P_R^(n)).
```

---

Even though it would seem that Algorithm 3 requires each player to know the other player's strategy, this is not the case and Algorithm 3 can be actually implemented in a distributed fashion with similar techniques as explained in [12].

## V. EXTENSION TO NON-SYMMETRIC CHANNELS

In this section we show how the tools developed for energy efficiency maximization in the symmetric case can be applied to asymmetric scenarios, too. The main difficulty is that obtaining closed-form sum rate expressions for non-symmetric channels is quite challenging. Moreover, even if some closed-form expressions could be obtained, they would be very cumbersome and involved, thus not allowing to gain any analytical insight, as we did in the symmetric case. Lacking an expression for the sum rate, we develop an alternative, yet equivalent, approach, which allows to formulate the GEE maximization problem based on the system rate region instead of the sum rate expression.[7]

Specifically, the GEE optimization problem from Section IV can be equivalently formulated as

$$\begin{cases} \max_{\substack{R_1,R_2,R_3,\\P_1,P_2,P_3,P_R}} & \dfrac{R_1 + R_2 + R_3}{\phi_1 P_1 + \phi_2 P_2 + \phi_3 P_3 + \psi P_R + P_c} \\ \text{s.t.} & (R_1, R_2, R_3) \in \mathcal{R}(P_1,P_2,P_3,P_R),\ P_R \in [0,\ P_R^{max}],\ P_i \in [0,\ P_i^{max}],\ i=1,2,3 \end{cases}, \qquad (27)$$

with $(R_1, R_2, R_3) \in \mathcal{R}(P_1,P_2,P_3,P_R)$ being the constraints which define the system rate region. Problem (27) can be seen to have a pseudo-linear objective, and therefore can be solved by means of Dinkelbach's algorithm as done in Section IV for the symmetric scenario. The complexity in solving (27) lies in the constraints $(R_1, R_2, R_3) \in \mathcal{R}(P_1, P_2, P_3, P_R)$, which are usually non-convex. Then, we have again a similar trade-off as in the symmetric scenario: low-complexity solutions can be obtained by means of sub-optimal methods, such as the alternating optimization, whereas the global solution can be obtained by means of monotonic optimization, at the expense of computational complexity.

---

[7]A similar approach has been used for sum rate optimization in the downlink of a MIMO multi-cell network [32].

As an illustrative example, consider AF-SND. It can be easily seen from the proof of Lemma 3 that $\mathcal{R}$ contains all non-negative rate tuples $(R_1, R_2, R_3)$ that satisfy

$$\sum_{i \in \mathcal{S}} R_i \leq \mathrm{C}\left(\frac{P_R \sum_{i \in \mathcal{S}} P_i}{N_R P_R + N_{q(k)}\left(N_R + \sum_{j \in \mathcal{K}} P_j\right)}\right), \tag{28}$$

for all $\mathcal{S} \in \{\{k\}, \{k, l(k)\} : k \in \mathcal{K}\}$. The RHS of (28) can be written as

$$\log\left(P_R \sum_{i \in \mathcal{S}} P_i + N_R P_R + N_{q(k)}\left(N_R + \sum_{j \in \mathcal{K}} P_j\right)\right) - \log\left(N_R P_R + N_{q(k)}\left(N_R + \sum_{j \in \mathcal{K}} P_j\right)\right),$$

which is a d.i. function, and therefore can be reformulated into a monotone constraint as done in Section IV for the symmetric scenario. Thus, monotonic optimization provides the global solution of (27). Instead, low-complexity solutions can be determined again by means of alternating optimization, alternatively optimizing the rate tuple $R_1, R_2, R_3$, the users' powers $P_1, P_2, P_3$, and the relay power $P_R$. It can be seen that each resulting sub-problem is a pseudo-linear maximization subject to convex constraints.

Similar considerations can be made for the other considered relaying protocols, too. As for AF-IAN, the asymmetric rate region can be straightforwardly obtained by following the proof for the symmetric case, which yields a formally equivalent expression as (3).

As for DF the achievable rate region in the asymmetric case is already available in [3], for the case of full message exchange. It can be seen that plugging the inequalities for the DF region of [3] into (27), yields again a problem which can be globally solved by monotonic optimization, whereas low-complexity solutions can be obtained by alternating optimization. Extending the approach used in [3] for the full message exchange to our considered partial message exchange is not difficult, and yields similar inequalities as in [3], which therefore can be also handled by our approach.

Finally, as for NNC, the rate region in the asymmetric scenario can be obtained by following the arguments in Section III. However, lengthier, but not difficult, derivations are required, since one has to include $Q_k$, $k \in \mathcal{K}$, as variables into the resource allocation problem. We omit further details here, being our focus on the resource optimization problem. It suffices to remark that also in this case we obtain expressions which when plugged into (27) yield a problem which can be globally solved by means of monotonic optimization, while low-complexity solutions can be obtained by means of sub-optimal approaches such as alternating optimization.

## VI. NUMERICAL RESULTS

For a discussion and numerical evaluation of the presented transmission schemes, we consider completely symmetric channels with $N = N_S = N_R$ and $P = P_S = P_R$. For the EE evaluation, we have $P^{max} = P_S^{max} = P_R^{max}$ and define $\text{SNR}^{\max} = P^{max}/N$. At first, we will assume $P_c = 1\,\text{W}$ and no power loss at the transmitter, i.e., $\psi = 1$ and $\phi = 3$, to compare algorithms and gain some general insight on the EE. In the next subsection, we model a wireless $200\,\text{GHz}$ board-to-board communication system to obtain more realistic simulation parameters.

Fig. 3 shows the achievable sum rates from Section III as a function of the SNR. It reflects the analytical results derived in Section III-E. First, observe that both NNC and AF-SND scale with the outer bound in the high SNR regime. Second, DF achieves the sum capacity up to a SNR of $8.1\,\text{dB}$ and is the best performing scheme up to $14.27\,\text{dB}$ (see Remarks 4 and 6). In the high SNR regime, its gap to the outer bound grows unbounded. Finally, AF-IAN scales with DF in the high SNR regime. It can be shown using the same methods as in Section III-E that its gap to DF approaches exactly $2\,\text{bit}$ as $\text{SNR} \to \infty$.

Fig. 4 illustrates the performance of the cooperative GEE maximization algorithms developed in Section IV-A. As for the $\text{GEE}_2$-type optimization problems, the performance of both monotonic optimization and alternating optimization are reported and it is seen that they perform virtually equally, thus suggesting that alternating optimization achieves the global optimum of the GEE. This is also the case in all other numerical simulations carried out in this paper. Thus, the alternating optimization algorithm is a good alternative to monotonic optimization when low computational complexity is the primary concern.

Fig. 4a shows the GEE as a function of the SNR for a fixed circuit power $P_c = 1\,\text{W}$ and unit noise variance. First of all, it can be seen that the GEE saturates when $\text{SNR}^{\max}$ exceeds a given value, which is lower than $5\,\text{dB}$ for all considered schemes. This is explained recalling that, unlike the achievable rate, the GEE is not increasing with the transmit powers, but instead admits an optimum transmit power level. If $P^{max}$ is larger than such power level, then it is not optimal to transmit at full power. This also explains why DF performs significantly better than all other schemes, including NNC. Indeed, due to the saturation of the GEE, the SNR range for which NNC yields a larger achievable sum rate than DF is not reached when GEE is optimized. Finally, as expected, NNC is better than AF-SND, which is better than AF-IAN.

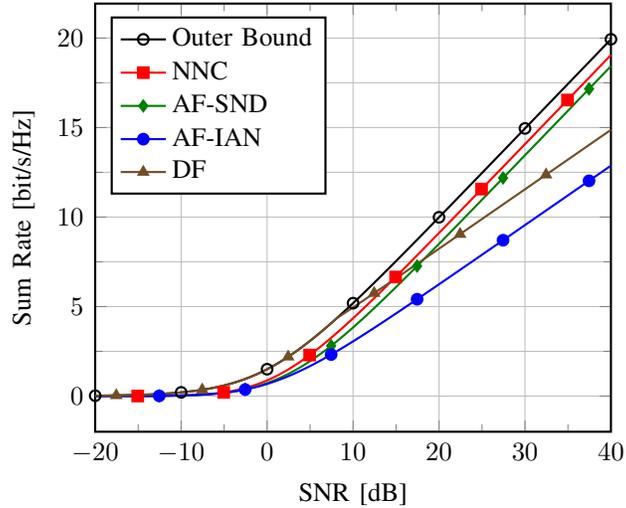

Fig. 3. Achievable sum rates in the 3-user MWRC; 1) Outer bound from Lemma 1, 2) NNC, 3) AF-SND and 4) AF-IAN, and 5) DF plotted as a function of the SNR.

In Fig. 4b the same simulation parameters as in Fig. 4a have been used but with different noise variance $N = 0.1\,\mathrm{mW}$. This results in a much higher saturation point which is approximately 30 dB and produces a different behavior of the curves. In the low SNR regime, DF achieves the outer bound, while in the high SNR regime NNC and AF-SND perform best while DF has significantly lower GEE. Due to the different DoFs, it can be expected that for even higher saturation points the gap between NNC and DF grows further, while the gaps between NNC and AF-SND, and between DF and AF-IAN stay nearly the same.

Next, Figs. 5 and 6 consider the competitive scenario of Section IV-B, also in comparison with the performance of the cooperative algorithms. The same simulation parameters as in Fig. 4a have been used, and it is assumed that every node consumes the same circuit power. Thus, $P_{c,S} = \frac{3}{4}P_c$ and $P_{c,R} = \frac{1}{4}P_c$. In particular, Fig. 5 compares the GEE achieved by the competitive power control algorithms to that obtained by the cooperative allocation of Section IV. It is seen that competitive and cooperative power control perform virtually the same if DF is used, whereas a limited gap appears if the other relaying strategies are employed. This can be intuitively explained recalling that a more significant coupling in a non-cooperative game results in a higher price of anarchy, and in our scenario, the game for DF is more coupled than the game for the other schemes. Indeed, the numerator of the players EEs for DF is given by the equation on the left in (20), while for the other schemes it is given by the equation on the right. In the latter case, the two variables $P_S$ and $P_R$ are more heavily coupled than in the former one because they appear in the same $\text{C}(\cdot)$ expression.

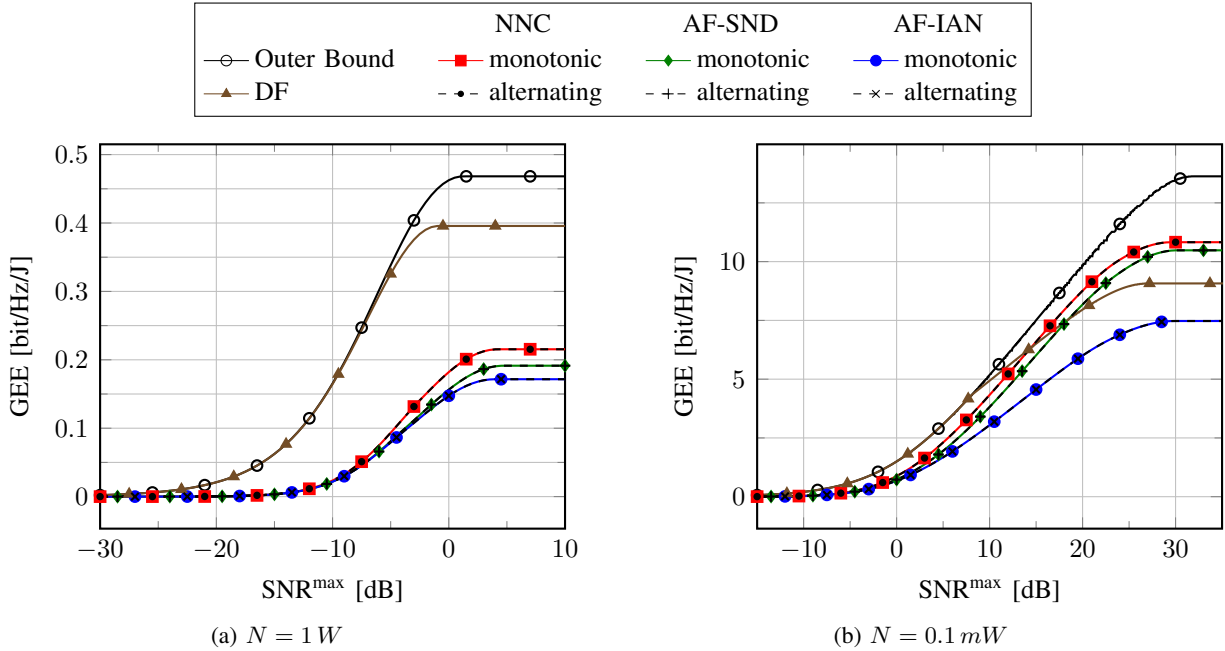

Fig. 4. GEE in the 3-user MWRC with 1) NNC, 2) AF-SND and 3) AF-IAN, 4) DF, and 5) the outer bound from Lemma 1 as a function of the SNR for fixed circuit power $P_c = 1\,\text{W}$ and noise variances $N = 1\,\text{W}$ and $N = 0.1\,\text{mW}$ in the left and right figure, respectively. The GEEs of NNC and AF are reported for both alternating and monotonic optimization.

Instead, Fig. 6 shows the behavior of the competitive algorithms for DF and NNC for different initialization values. It is observed that DF is more sensitive to the initialization point than NNC. In particular, the closer the initialization point to $P^{max}$, the sooner the GEE reaches the saturation level. This can be explained recalling that for DF the BRs are expressed as in (21), where the variables $\bar{P}_S^{max}$ and $\bar{P}_R^{max}$ appear, which are increasing in the strategy of the other player. If the initialization values of the transmit power is low with respect to $P^{max}$, the corresponding values $\bar{P}_S^{max}$ and $\bar{P}_R^{max}$ will also be low and will likely prevail in the $\min$ operation, preventing from transmitting at the saturation levels $\bar{P}_S$ and $\bar{P}_R$. This does not occur for NNC and AF since the BR has a different expression which does not depend on similar variables as $\bar{P}_S^{max}$ and $\bar{P}_R^{max}$.

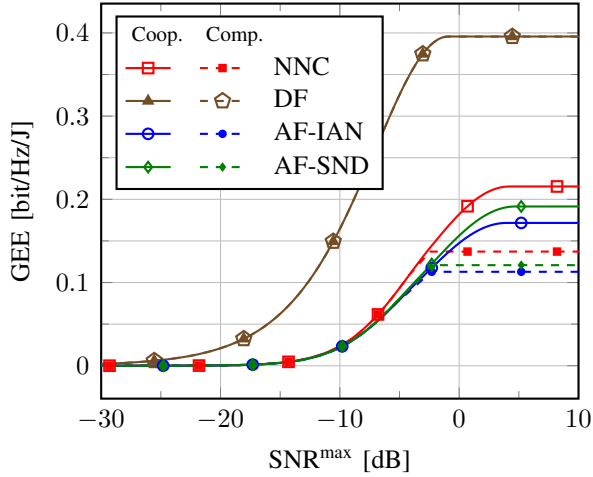

Fig. 5. Cooperative vs. competitive energy efficiency maximization for fixed circuit powers $P_{c,S} = 750\,\text{mW}$ and $P_{c,R} = 250\,\text{mW}$, and noise variance $N = 1\,\text{W}$.

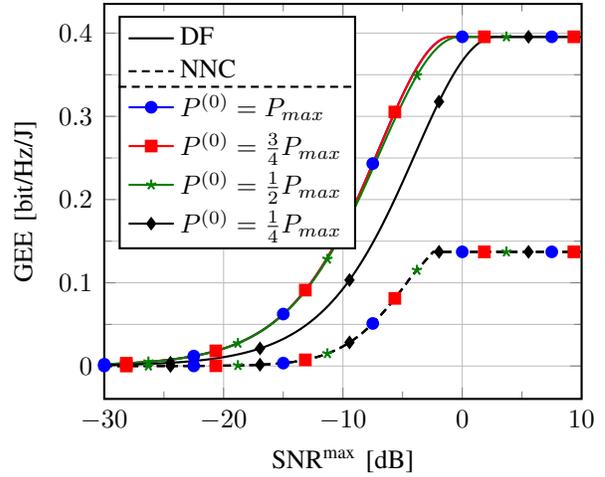

Fig. 6. Competitive energy efficiency maximization with different initializations. The same simulation parameters as in Fig. 5 have been used.

Comparing previous illustrations and especially Fig. 4, we learn that the energy-efficient performance of the different transmission and relaying scheme highly depends on the saturation point of the GEE function, which in turn is heavily affected by the the system parameters. Thus, for a sensible comparison of the different schemes, it is necessary to choose physically meaningful values for the noise variance and the circuit power consumption. Furthermore, assuming the same circuit power consumption for all schemes is unfair since they have different hardware complexity. For this reason, in the following subsection we consider a specific communication system and develop a simple model for its power consumption which allows for a fair comparison of the considered schemes.

### A. mmWave Board-to-Board Communications

We consider a wireless board-to-board communication system operating at $200\,\text{GHz}$ carrier frequency with $25\,\text{GHz}$ bandwidth. Channel measurements and a link budget for such a system with 4-by-4 antenna arrays are given in [2]. Since there is no mobility in this scenario, the beams have a fixed direction and the antenna array combined with the beamforming network can be regarded as a single antenna with a gain of $7\,\text{dB}$. With a link length of $0.1\,\text{m}$, the resulting channel gain is -65.8 dB as reported in [2, Table I].[8,9] As for the noise, we have $N_S = N_R = k_B T_c$ where $k_B$ is the Boltzmann constant and $T_c$ the absolute temperature of operation.

Next, we develop a model for the circuit power consumption $P_c$ that reflects the different hardware complexities of the considered transmission schemes. We assume that $P_c$ depends neither on the power allocation nor on the transmission rate for simplicity. The analog part of each receiver consists of an Rx frontend and two analog-to-digital converters (ADCs) consuming a total power of $P_{ADC}$. Similarly, each transmitter consists of a Tx frontend, two digital-to-analog converters (DACs) (consuming a power $P_{DAC}$), and a power amplifier (PA) with efficiency $\eta$, i.e., $\psi = \frac{1}{\eta}$ and $\phi = \frac{3}{\eta}$. The Rx frontend has 16 low noise amplifiers (LNAs), one for each antenna, two mixers and one LO-driver (local oscillator) for those mixers resulting in a power $P_{rx}$. Similarly, the Tx frontend consists of two mixers and a LO-driver consuming a power $P_{tx}$.[10] Realistic numbers for hardware-dissipated power and PA efficiencies are reported in Table II, and the resulting model parameters are given in Table III. We define $P_{c,analog} = P_{rx} + P_{ADC} + P_{DAC} + P_{tx}$. The digital signal processing (DSP) power consumption $P_{DSP}$ of each node is modeled as a multiple of the power $P_{dec}$ required by a single user decoder. From the LDPC decoder model in [33][11] we get a decoding power of 267.6 mW. Adding some power to account for other processing such as encoding, we round $P_{dec}$ to 300 mW.

As for the users, when interference is treated as noise, only one message needs to be decoded, hence $P_{c,S}^{IAN} = P_{c,analog} + P_{dec}$. For SND, we model the power consumption as if the two messages were decoded sequentially, hence, $P_{DSP} = 2P_{dec}$.[13] A similar assumption is used for receive decoding if DF relaying is used because also in this case the receiver has to decode two messages. Thus, $P_{c,S}^{SND} = P_{c,S}^{DF} = P_{c,analog} + 2P_{dec}$.

As for the relay, DF uses a 3-user MAC decoder. Hence, $P_{c,R}^{DF} = P_{c,analog} + 3P_{dec}$. The NNC relay does not decode the received messages, and, in our scenario, does not have a message to transmit on its own. Instead it uses a vector quantizer to compress the received sequence. Following [37], we assume that the (scalar) ADC in our model roughly models the quantization operation of NNC. The only DSP operation then is encoding, and we have $P_{c,R} = P_{c,analog} + 0.1 P_{dec}$.[14] In contrast to the other relaying schemes, AF directly amplifies the analog signal, so neither DSP nor ADC and DAC are necessary. Thus, $P_{c,R}^{AF} = P_{rx} + P_{tx}$. Finally, the circuit power $P_c = 3P_{c,S} + P_{c,R}$.

---

[8]Computed without considering the polarization mismatch since we do not consider polarization here.

[9]In previous sections, unit channel gain and bandwidth have been assumed in the sum rate expressions. However, previous results and algorithms can be straightforwardly extended to the case in which the powers $P_S$ and $P_R$ in $R_\Sigma$ are scaled by a channel coefficient and divided by the bandwidth.

[10]We do not include the PA in the Tx frontend since it is modelled separately.

[11]For our model, the following parameters are suitable: $E_{edge} = 10^8 \cdot k_B \cdot T$, $r_{COD} = \frac{1}{2}$, $R = 50\,\text{GBit/s}$, $\lambda = 3$, $l = 2$.

[12]Those values have been provided by project partners from the DFG CRC 912 "Highly Adaptive Energy-Efficient Computing".

[13]Recall that NNC uses SND at the receivers.

TABLE II
EXEMPLARY POWER CONSUMPTIONS FOR 200 GHZ RF COMPONENTS.

| Component | Value | Unit | Reference |
|---|---|---|---|
| Mixer | 17 | $mW$ | Unpublished measurements[12] |
| LO-driver | 24 | $mW$ | |
| LNA | 18 | $mW$ | [34] |
| ADC | 406 | $mW$ | [35] |
| DAC | 400 | $mW$ | Estimate[12] |
| PA efficiency | 6.2 | % | [36] |

TABLE III
PARAMETERS FOR CIRCUIT POWER MODELLING OF BOARD-TO-BOARD COMMUNICATIONS.

| Component | Symbol | Value | Unit |
|---|---|---|---|
| Rx Frontend | $P_{rx}$ | 346 | $mW$ |
| Analog-to-digital converter | $P_{ADC}$ | 812 | $mW$ |
| Single User Decoder | $P_{dec}$ | 300 | $mW$ |
| Digital-to-analog converter | $P_{DAC}$ | 800 | $mW$ |
| Tx Frontend | $P_{tx}$ | 58 | $mW$ |
| Power amplifier | $\eta$ | 6.2 | % |

Fig. 7 was obtained using the developed power consumption model with the algorithms from Section IV-A. It can be seen that in the low SNR regime DF performs best but that starting from approximately 10 dB AF-SND achieves the best GEE. This continues to hold for higher SNR's, where DF is outperformed by all other schemes. Since there are currently no NNC implementations available, we might have underestimated its power consumption. Let us assume its decoding complexity is actually higher than that of DF and add an additional $P_{dec}$ for each user and set $P_{DSP} = 4P_{dec}$ for the relay (instead of $0.1P_{dec}$). The resulting GEE is shown as a dashed line in Fig. 7. It can be seen that it still outperforms DF in the high SNR regime despite its much higher power consumption.

## VII. CONCLUSION

In this paper, we studied both achievable sum rates and the EE of the symmetric 3-user MWRC with a partial message exchange. We provided analytic sum rate expressions for the most common relaying schemes and characterized the sum capacity for certain SNR regimes. Next, we discussed the problem of energy-efficient power control both in the cooperative and competitive scenario. The results indicate that competitive optimization suffers a limited gap with respect to cooperative resource allocation, and that the proposed low-complexity cooperative allocation achieves the global optimum.

---

[14]We assume $P_{DSP} = 0.1P_{dec}$, which is roughly the power we accounted for *other processing* in $P_{dec}$ before.

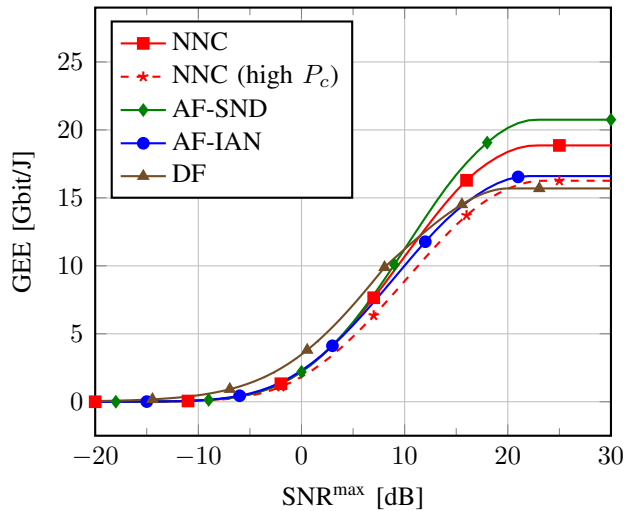

Fig. 7. GEE in the 3-user MWRC of 1) NNC, 2) AF-SND and 3) AF-IAN, 4) DF, and 5) the outer bound from Lemma 1 as a function of the SNR for wireless board-to-board communication at $200\,\text{GHz}$ as discussed in Section VI-A.

Moreover with reference to a realistic mmWave board-to-board communication system, we have shown that AF performs best for SNRs above $10\,\text{dB}$, closely followed by NNC, while DF suffers from a higher hardware power consumption. However, as discussed in Section VI, the performance of the different scheme is significantly affected by the simulation parameters. Further analysis is required to understand which is the most energy-efficient relaying scheme in other communication systems.

## Acknowledgement

The authors would like to thank C. Carta, G. Tretter, D. Fritsche, and N. ul Hassan from Technische Universität Dresden for valuable discussions on the circuit power consumption.